# Phylogeny of Twenty-One Mammals

## Ray Han

**Introduction**

Phylogeny can be inferred using two sources of data from an organism: morphological data and molecular data. Historically, phylogenies were usually inferred using morphological characters (Lee and Palci, 2015), but some morphological features may not necessarily indicate shared heritage (Zou and Zhang, 2016). With the introduction of molecular phylogenies, the base sequence of genes, or amino acid sequence of proteins can be compared to find the number of similarities or differences to ascertain levels of relatedness between species. These two types of phylogenies are to be taken as a data-driven hypothesis about the evolutionary history of the studied organisms, and a consensus is drawn from the comparison between the different phylogenies built from the two sources of data, utilizing different methods.

It has been suggested that molecular phylogenies are more accurate than morphological phylogenies because it is believed that molecular characters are less susceptible to convergent evolution (Zou and Zhang, 2016). However, this does not mean that molecular phylogenies are to be considered over morphological phylogenies without indication to do so (Pisani et al., 2007). In some cases, morphological phylogenies can be the only phylogenies created, such as when the organism is extinct, and molecular data is unable to be gathered (Zou and Zhang, 2016). Therefore, using both morphological phylogenies and molecular phylogenies can provide independent evidence to support a hypothesis (Bucklin and Frost, 2009).

Two studies that related whales to taxa in the order Artiodactyla are good examples of using morphological phylogenies and molecular phylogenies to provide a unique view to support a hypothesis. The first study used a molecular approach, by characterizing two families of non-coding DNA sequences. Their findings resolved the phylogenetic relationships between whales, ruminants, hippopotamuses, and pigs, and provided evidence to support the hypothesis that whales, ruminants, and hippopotamuses formed a monophyletic group. Their findings challenged the traditional view that the order Artiodactyla, which included even-toed ungulates, is monophyletic (Shimamura et al., 1997). Their findings were consistent with an earlier study, that used a morphological approach, comparing the morphological traits of whales and Basilosaurus, and found that the feet were paraxonic, i.e., bearing weight in 2 parallel axes within the foot in the order Cetacea which includes whales, which was consistent with serological evidence of a relationship to the order Artiodactyla (Gingrich et al., 1990).

This paper's goal is to infer the evolutionary history of twenty-one mammals by comparing a morphological phylogeny to a molecular phylogeny made by using the cytochrome c oxidase subunit 1 (COX-1) protein. From the literature, it is expected that the molecular and morphological phylogenies will have many clades overlapping, with minor differences (Pisani et al., 2007).

**Materials and methods**

*Proteomic phylogeny* - Cytochrome c oxidase subunit 1 (COX-1) protein is a component of cytochrome c oxidase, the last enzyme in the mitochondrial electron transport chain, that helps facilitate oxidative phosphorylation inside the mitochondria. The gene that encodes this protein is among the most conserved mitochondrial protein-coding genes in animals (Folmer et al., 1994), meaning its sequence remains relatively unchanged through evolution. This makes it a good protein to compare for evolutionary studies (Miya and Nishida, 2000) and is commonly used in research about animals (Girard et al., 2022).

To compare the twenty-one mammal taxa (Table 1), the amino acid sequences were obtained (The UniProt Consortium, 2023). These sequences were aligned by Clustal Omega (Sievers et al., 2011; McWilliam et al., 2013), which uses a method to align sequences using mathematical models and identifies areas within the sequences that are similar (Sievers et al., 2011; McWilliam et al., 2013). The percent identity matrix generated by Clustal Omega was saved (Table 2), and the alignment results were also saved in FASTA format.

The computer software Molecular Evolutionary Genetic Analysis (Tamura et al., 2021; Stecher et al., 2020) was used to construct the proteomic phylogenetic tree, using the maximum likelihood statistical method. Using this, an initial tree is built using a fast but suboptimal method. The branch lengths of this initial tree are then adjusted for maximum likelihood for the data set using a model of evolution, and variants of the branching pattern of the tree are then created using different methods to search for branching patterns that fit the data better. Branch lengths are then computed for these variant trees and the greatest likelihood is retained as the best choice so far, and this search continues until no greater likelihoods are found. Some settings were set within the tree window for the initial tree method, model of evolution, and branching pattern variant methods. These settings were all on default settings, except for the branching pattern variant method, which was set as extensive subtree-pruning-regrafting. Subtree-pruning-regrafting prunes parts of the initial tree and then rearranges them, and then calculates the likelihood. This is done until the likelihoods of all possible tree orientations are calculated, and the branching pattern with the highest likelihood value is deemed the best tree (Wu, 2009). This was chosen because it provided a more expansive search for tree branching patterns than the other available option (Zhou et al., 2017).

*Morphological Phylogeny* - A morphological phylogeny was produced and obtained from the morphological traits of the mammals. It was made to be most parsimonious.

*Comparison* - The taxa were compared by clades, and which taxa were placed differently. Similar and identical clades between the proteomic phylogenies and the morphological phylogenies were noted and highlighted. Major differences in taxa placement between the phylogenies were also identified. The clades in the phylogenies were summarized in a table (Table 3), with taxa higher on the column being more closely related within the clade (e.g. the top two taxa are sister taxa), and differences highlighted in color.

**Results**

After comparing the morphological phylogeny (Figure 1), and the proteomic phylogeny (Figure 2), it was noted that the phylogenies mostly had similar clades, with some minor differences, as summarized in Table 3. Out of twenty-one taxa, fifteen taxa were grouped into the labeled clades I, II, III, IV, and V, which were present in and marked in both of the phylogenies, with the sister taxa arrangement between the taxa in clades II, III, IV, and V being identical. Six taxa were not grouped into clades, with five being identical, being *Colobus polykomos* (Colobus Monkey), *Cavia porcellus* (Guinea Pig), *Rattus norvegicus* (Rat), *Oryctolagus cuniculus* (Rabbit), and *Ornithorhynchus anatinus* (Platypus). The non-identical taxa were *Tamandua tetradactyla* (Anteater) in the morphological phylogeny and the *Equus caballus* (Horse) in the proteomic phylogeny. It was also noted that clade I had two versions, clade $I_1$ in the morphological phylogeny, and clade $I_2$ in the proteomic phylogeny.

As shown in Table 3, clade I had two versions conflicting between both of them. Clade $I_1$ included *E. caballus*, *Dendrohyrax dorsalis* (Tree hyrax), *Loxodonta africana* (Elephant), *Trichechus manatus* (Manatee), and *E. caballus*, but in clade $I_2$, *E. caballus* was not present and was replaced with *T. tetradactyla*. In the morphological phylogeny, *T. tetradactyla* was more closely related to the taxa in clade IV, and in the proteomic phylogeny, *E. caballus* was more closely related to clade II. Also, in clade $I_1$, *D. dorsalis,* and *L. africana* were sister taxa, while in clade $I_2$, *D. dorsalis,* and *T. manatus* were the sister taxa. Through comparison of the proteomic phylogeny (Figure 2) and morphological phylogeny (Figure 1), it was observed that although not forming a clade in the morphological phylogeny, as they do in the proteomic phylogeny, *C. porcellus* (Guinea pig), and *O. cuniculus* were closely related in both phylogenies.

A major difference between the proteomic and morphological phylogenies was that in the proteomic phylogeny, *C. polykomos* was the most basal lineage, and *O. anatinus* was nested in other taxa, but the opposite was seen in the morphological phylogeny, with *O. anatinus* being the most basal lineage, while *C. polykomos* was nested in other taxa.

It was also observed that an even amount of toes, carnassial teeth, and the presence of a marsupium during gestation were associated with a clade that was also present in the proteomic phylogeny. Also observed was that aquatic taxa were mostly more closely related to terrestrial taxa than other aquatic taxa.

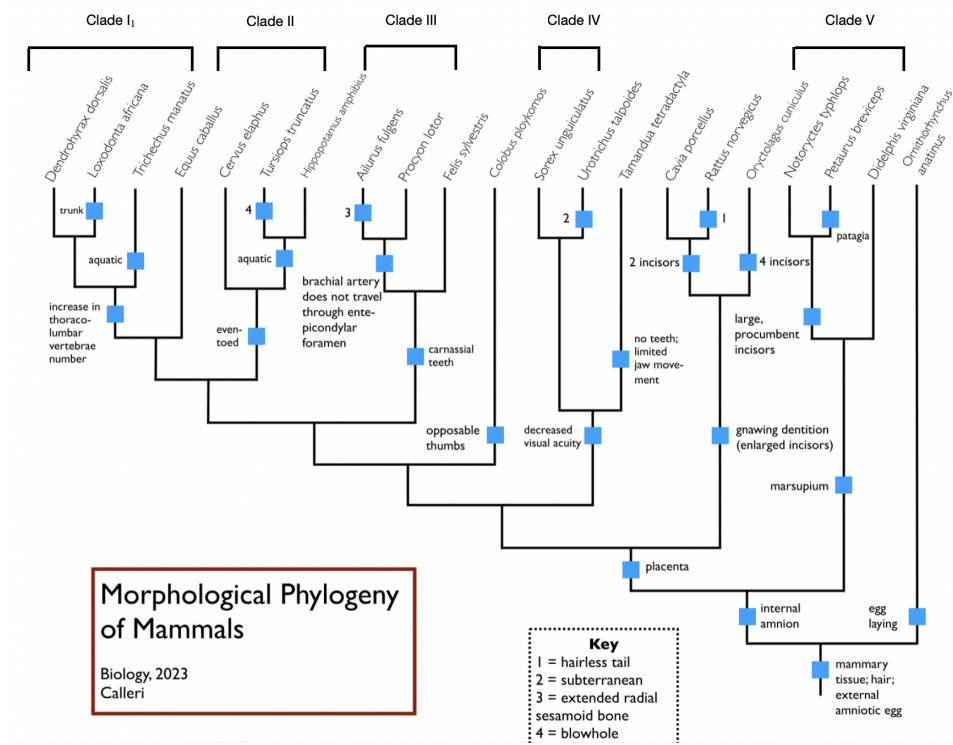

*Figure 1*: *Morphological phylogeny of the twenty-one mammals, with common clades marked.*

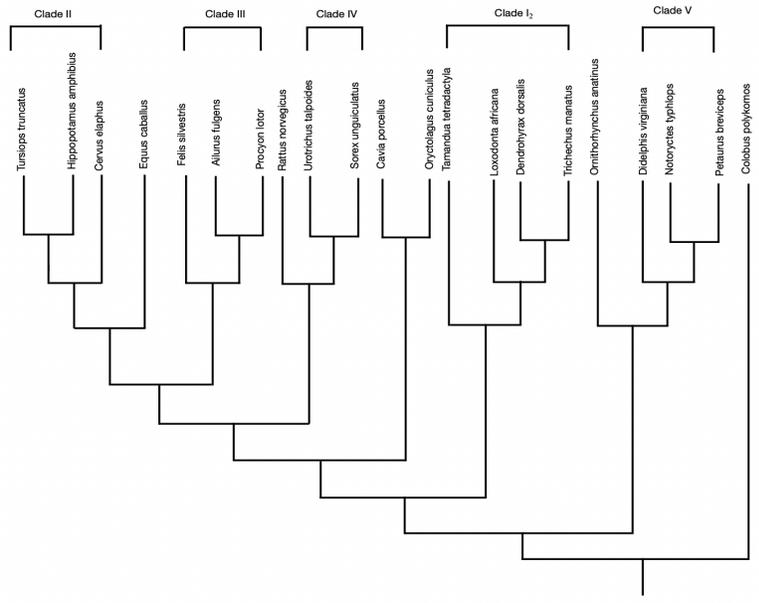

***Figure 2**: Proteomic phylogeny of twenty-one mammals, with common clades marked.*

# Mammal Phylogeny Taxa List

| Common Name | Scientific Name | CO1 (UniProt) | CO1 (NCBI ID) |
|---|---|---|---|
| Hippo | *Hippopotamus amphibius* | Q9ZZY9 | 9833 |
| Horse | *Equus caballus* | P48659 | 9796 |
| Dolphin | *Tursiops truncatus* | B9UD78 | 9739 |
| House Cat | *Felis silvestris* | P48888 | 9685 |
| Red Deer | *Cervus elaphus* | Q2V097 | 9860 |
| Guinea Pig | *Cavia porcellus* | Q9TEG9 | 10141 |
| Rat | *Rattus norvegicus* | P05503 | 10116 |
| Rabbit | *Oryctolagus cuniculus* | O79429 | 9986 |
| Colobus Monkey | *Colobus polykomos* | O99041 | 9572 |
| Racoon | *Procyon lotor* | Q3L6R3 | 9654 |
| Manatee | *Trichechus manatus* | B0JDY8 | 9778 |
| Anteater | *Tamandua tetradactyla* | Q8LWP0 | 48850 |
| Opossum | *Didelphis virginiana* | P41310 | 9267 |
| Sugar Glider | *Petaurus breviceps* | Q1MWH1 | 34899 |
| Tree Hyrax | *Dendrohyrax dorsalis* | B0JDW6 | 42325 |
| Elephant | *Loxodonta africana* | Q9TA27 | 9785 |
| Shrew | *Sorex unguiculatus* | Q94YE2 | 62275 |
| Red Panda | *Ailurus fulgens* | A7DXW2 | 9649 |
| Platypus | *Ornithorhynchus anatinus* | Q36452 | 9258 |
| Marsupial Mole | *Notoryctes typhlops* | Q5QRZ9 | 37699 |
| Mole | *Urotrichus talpoides* | Q7Y8E0 | 106106 |

*Table 1*: List of twenty-one mammals with common names, scientific names, and protein IDs.

| | Colobus | Marsupial Mole | Sugar Glider | Opossum | Mole | Shrew | Platypus | Elephant | Rat | Tree Hyrax | Dolphin | Guinea Pig | Manatee | Ant eater | Rabbit | Hippo | Red Panda | Raccoon | Cat | Red deer | Horse |
|---|---|---|---|---|---|---|---|---|---|---|---|---|---|---|---|---|---|---|---|---|---|
| Colobus | | | | | | | | | | | | | | | | | | | | | |
| Marsupial Mole | 88.5 | | | | | | | | | | | | | | | | | | | | |
| Sugar Glider | 88.89 | 96.3 | | | | | | | | | | | | | | | | | | | |
| Opossum | 90.06 | 96.49 | 96.69 | | | | | | | | | | | | | | | | | | |
| Mole | 88.13 | 93.96 | 93.76 | 93.57 | | | | | | | | | | | | | | | | | |
| Shrew | 87.74 | 93.37 | 92.79 | 92.79 | 96.11 | | | | | | | | | | | | | | | | |
| Platypus | 87.72 | 92.98 | 92.4 | 92.2 | 91.81 | 92.2 | | | | | | | | | | | | | | | |
| Elephant | 87.94 | 91.42 | 90.84 | 91.03 | 91.25 | 91.63 | 90.06 | | | | | | | | | | | | | | |
| Rat | 89.69 | 91.81 | 91.42 | 91.23 | 92.8 | 93.39 | 91.81 | 90.66 | | | | | | | | | | | | | |
| Tree Hyrax | 88.72 | 93.18 | 92.59 | 92.79 | 92.8 | 91.63 | 90.64 | 91.83 | 92.22 | | | | | | | | | | | | |
| Dolphin | 87.16 | 91.42 | 91.42 | 91.62 | 92.02 | 90.86 | 90.25 | 90.29 | 91.83 | 92.22 | | | | | | | | | | | |
| Guinea Pig | 90.45 | 92.79 | 92.4 | 93.57 | 93.57 | 92.79 | 92.79 | 94.35 | 93.57 | 92.4 | | | | | | | | | | | |
| Manatee | 89.67 | 92.79 | 93.18 | 93.76 | 93.76 | 93.18 | 92.2 | 93.76 | 93.18 | 95.32 | 93.96 | 94.54 | | | | | | | | | |
| Ant eater | 89.67 | 93.18 | 92.79 | 93.39 | 93.57 | 93.18 | 92.59 | 92.98 | 93.57 | 94.74 | 92.98 | 94.93 | 96.1 | | | | | | | | |
| Rabbit | 90.84 | 93.18 | 93.18 | 93.57 | 93.96 | 93.37 | 93.37 | 93.37 | 94.15 | 94.54 | 92.79 | 96.3 | 95.52 | 96.3 | | | | | | | |
| Hippo | 89.11 | 92.98 | 92.59 | 93.18 | 93.58 | 92.8 | 91.62 | 92.8 | 92.8 | 94.75 | 94.95 | 94.15 | 94.93 | 94.74 | 94.74 | | | | | | |
| Red Panda | 90.08 | 93.76 | 93.96 | 93.76 | 95.14 | 93.39 | 92.59 | 92.8 | 94.55 | 93.77 | 93.58 | 95.52 | 95.32 | 95.52 | 96.3 | 94.94 | | | | | |
| Raccoon | 89.3 | 93.37 | 93.18 | 93.18 | 94.16 | 93 | 91.81 | 92.41 | 93.97 | 93.39 | 93.19 | 94.93 | 94.54 | 94.54 | 95.52 | 94.55 | 98.44 | | | | |
| Cat | 90.08 | 93.57 | 93.76 | 93.37 | 95.33 | 93.77 | 92.79 | 93 | 93.97 | 94.55 | 94.16 | 95.13 | 95.91 | 95.32 | 96.49 | 96.11 | 97.08 | 97.47 | | | |
| Red deer | 89.49 | 94.15 | 94.15 | 93.76 | 95.33 | 93.58 | 92.4 | 93.39 | 93.97 | 94.94 | 96.11 | 94.93 | 96.3 | 96.1 | 95.91 | 97.08 | 96.89 | 96.3 | 97.47 | | |
| Horse | 89.88 | 93.37 | 93.96 | 93.37 | 95.33 | 93.39 | 92.59 | 92.61 | 94.16 | 94.36 | 95.14 | 95.13 | 96.1 | 95.32 | 96.1 | 95.53 | 97.28 | 96.69 | 97.67 | 98.05 | |

*Table 2*: Percent identity matrix generated from Clustal Omega alignment.

| I | II | III | IV | V |
|---|---|---|---|---|
| **Morphologic Phylogeny** | | | | |
| Dendrohyrax dorsalis (Tree Hyrax) | Tursiops truncates (Dolphin) | Ailurus fulgens (Red Panda) | Sorex unguiculatus (Shrew) | Nororyctes typhlops (Marsupial Mole) |
| Loxodonta Africana (Elephant) | Hippopotamus amphibius (Hippo) | Procyon lotor (Racoon) | Urotrichus talpoides (Mole) | Petaurus breviceps (Sugar Glider) |
| Trichechus manatus (Manatee) | Cervus elaphus (Red Deer) | Felis silvestris (House Cat) | | Didelphis virginiana (Opossum) |
| Equus caballus (Horse) | | | | |
| **Molecular Phylogeny** | | | | |
| Dendrohyrax dorsalis (Tree Hyrax) | Tursiops truncatus (Dolphin) | Ailurus fulgens (Red Panda) | Urotrichus talpoides (Mole) | Nororyctes typhlops (Marsupial Mole) |
| Trichechus manatus (Manatee) | Hippopotamus amphibius (Hippo) | Procyon lotor (Racoon) | Sorex unguiculatus (Shrew) | Petaurus breviceps (Sugar Glider) |
| Loxodonta africana (Elephant) | Cervus elaphus (Red Deer) | Felis silvestris (House Cat) | | Didelphis virginiana (Opossum) |
| Tamandua tetradactyla (Anteater) | | | | |

*Table 3*: Table with the taxa in the clades, decreasing relatedness down the column, and differences highlighted.

**Discussion**

The goal of the paper was to form a consensus about the evolutionary histories of twenty-one mammals by comparing a morphological phylogeny created using morphological traits, and a molecular phylogeny created using the COX-1 protein. A comparison of these phylogenies showed large similarities between the proteomic and morphological phylogenies, with smaller differences. These differences could be the result of convergent evolution in the morphological traits, which is believed to happen more often than in molecular traits (Zou and Zhang, 2016), as well as different methods of processing data.

The presence of an even amount of toes, carnassial teeth, and the presence of a marsupium during gestation was associated with a clade that was also present in the proteomic phylogeny. Since each one of these defines a single clade, it can be hypothesized that these traits are affected by changes in the gene that encodes for the COX-1 protein. However, there is no such research currently to support this.

As previously mentioned, not all of the aquatic taxa were not closely related, and instead of being in a clade together, they were usually more closely related to other terrestrial taxa. These aquatic taxa are *T. manatus* in clade $I_1$, as well as *T. truncatus,* and *Hippopotamus amphibius* (Hippo) in clade II. It can be hypothesized that the aquatic taxa and the closely related terrestrial taxa had a common ancestor which was terrestrial, since there are many more terrestrial taxa, and eventually evolved to become aquatic. In the case of clade $I_1$, there is evidence that *L. africana*, *D. dorsalis,* and *T. manatus* were descended from a common ancestor (Prothero, 2009). In the case of clade II, this hypothesis is also supported by the findings that taxa of the order Cetacea, which includes *Tursiops truncatus* (Dolphin), are related to taxa of the order Artiodactyla, which includes *H. amphibius* and *Cervus elaphus* (Red deer) (Gingrich et al., 1990). It has also been well-documented that Cetaceans and *H. amphibius* had land ancestors (Thewissin et al., 2009; Orliac et al., 2023). This is also supported by the fact that it would have been more parsimonious to evolve the character of being aquatic than to lose the character, therefore it can be suggested that these aquatic mammals evolved from terrestrial ancestors.

It has been suggested that morphological phylogenies are less accurate than molecular ones. However, this is not to say that morphological phylogenies are to be completely disregarded. Instead, they are to be taken into account as much as molecular phylogenies when their results conflict and molecular phylogenies should not immediately be considered to be more reliable without evidence (Pisani et al., 2007). One major difference between the proteomic and morphological phylogenies was the most basal taxon, with the most basal taxon in the morphological phylogeny being *O. anatinus*, and the most basal taxon in the proteomic phylogeny being *C. polykomos*. In the case of the different basal taxa, Research showed that the subfamily Colibinae, which included *C. polykomos* diverged about 5-9 million years ago (van der Kuyl et al., 1994), and families of platypus, which included *O. anatinus,* diverged about 17-90 million years ago (Bino et al., 2019). From this evidence, it is proposed that *O. anatinus* is the most basal taxon within the phylogeny of the twenty-one mammals.

One difference observed between the morphological and molecular phylogenies was the conflict between the 2 versions of clade I, namely the morphological version, clade $I_1$, and the proteomic version, clade $I_2$. There has been little evidence to suggest that one version of the clade is more accurate than the other, however, it has been well-documented that *D. dorsalis, L. africana,* and *T. manatus* are distantly related to each other (Kellogg et al., 2007). To resolve this conflict between the clades, it is proposed that *E. caballus* be excluded from the morphological clade $I_1$, *T. tetradactyla* be excluded from the proteomic clade $I_2$, and clade I simplified only to include *D. dorsalis, L. africana, and T. manatus*. Future research should be done to fully resolve the sister taxa arrangement of *D. dorsalis, L. africana, and T. manatus,* as it is still unclear as to whether the resolved $I_1$ or the resolved $I_2$ is more accurate in terms of sister taxa placement.

In this paper, the comparison and combination of two phylogenies, molecular and morphological, has been shown to be an appropriate method to come to a consensus about the evolutionary histories of twenty-one mammals, as the morphological and molecular phylogenies independently demonstrated the presence of four identical clades and one similar clade. Future research into this field could be done to determine the relatedness between the morphological traits, and changes in the COX-1 gene, as well as to fully resolve the differences between the proteomic clade $I_2$, and the morphological clade $I_1$. Furthermore, more research could be done to apply the phylogeny of the twenty-one mammal taxa to the universal tree of life, by comparing them with more taxa.